\documentclass[article]{aa}
\usepackage{txfonts}
\usepackage{natbib}
\bibpunct{(}{)}{;}{a}{}{,} 
\usepackage{graphicx}

\begin{document}

\headnote{Research Note:}

\title{The Opacity of Spiral Galaxy Disks VII:\\
the accuracy of galaxy counts as an extinction probe
\thanks{Research support by NASA 
through grant number HST-AR-08360 from the Space Telescope Science 
Institute (STScI), the STScI Discretionary Fund (grant numbers 82206 and 82304) 
and the Kapteyn Astronomical Institute of the University of Groningen.}
}
\author{B. W. Holwerda \inst{1,2} \and R. A. Gonzalez \inst{3} 
\and Ronald J. Allen \inst{2} \and P. C. van der Kruit \inst{1}}

\offprints{B.W. Holwerda, \email{Holwerda@stsci.edu}}

\institute{Kapteyn Astronomical Institute, postbus 800, 
9700 AV Groningen, the Netherlands.
\and 
Space Telescope Science Institute, 3700 San Martin Drive, Baltimore, MD 21218
\and
Centro de Radiastronom\'{\i}a y Astrof\'{\i}sica, Universidad Nacional Aut\'{o}noma de M\'{e}xico, 58190 Morelia, Michoac\'{a}n, Mexico}

\date{12/04/2005 / 19/07/2005}

\titlerunning{Limits of HST for the ``Synthetic Fields Method''}
\authorrunning{Holwerda et al.}

\abstract{
The ``Synthetic Field Method'' (SFM) was introduced by \cite{Gonzalez98} 
to calibrate numbers of distant galaxies as a probe of extinction in a 
foreground spiral disk. \cite{Gonzalez03} studied the effect of the foreground 
disk on these numbers using simulations of current and future instruments 
for fields in the LMC, M31 and NGC 4536, a galaxy in Virgo. 
They concluded that: (1) the brighter centers of disks were unsuitable, (2) 
the granularity of the disk at a fixed surface brightness is the limiting factor 
in the detection of distant galaxies, and (3) the optimum distance for 
measurements would be that of the Virgo cluster for the current instruments 
on board HST. At this distance the foreground disk is smoothed with distance, 
improving detection of distant background galaxies.  
\cite{Holwerda05a} automated the SFM and \cite{Holwerda05b} applied it to 
a large set of WFPC2 fields. In this paper, the quality of the extinction 
measurement in these fields is compared to their distance, granularity, 
surface brightness and structure. 
The average surface brightness of the of a field is shown to directly influence 
the accuracy of the SFM. This restricts meaningful measurements to the disks 
of spiral galaxies. 
Large structures such as spiral arms have a similar effect.
The granularity or small scale structure in a field influences the detection of 
distant galaxies, limiting the SFM measurements in nearby disks. 
From the trends in the accuracy and maximum practical field-of-view 
considerations, the minimum and maximum distance for SFM application, 
approximately 5 and 35 Mpc respectively.
Using the same instrument and detection method, the relations with SFM 
parameters and field characteristics can be used to forgo the synthetic 
fields altogether. 
For the wealth of ACS fields becoming available in the archive, 
these relations can be used to select those fields based on expected SFM accuracy.

 }

\maketitle

\section{\label{secLimintro}Introduction}

The number of field galaxies seen through a nearby foreground galaxy has 
for a long time been recognized as a possible probe into the dust extinction 
in the foreground object, much in the way star counts are used in our own 
Galaxy. \cite{Hubble34} noted a drop of field galaxies at lower Galactic latitude, 
a fact that was later used by \cite{Burstein82} to map the Galactic extinction 
based on counts from \cite{Shane67}. 

As number counts are limited by statistics, a measurement over the largest 
practical solid angle is needed. This prompted several studies of the LMC 
and SMC (\cite{Shapley51}, \cite{Wesselink61a}, \cite{Hodge74}, 
\cite{MacGillivray75}, \cite{Gurwell90} and \cite{Dutra01}), the majority on 
photographic plates. The dust effects in other galaxies were characterised 
by \cite{Zaritsky94}, \cite{Lequeux95} and \cite{Cuillandre01}. 

However, the detection of field galaxies is not only affected by the absorption 
in the foreground disk. The crowding and confusion of the foreground disk 
also play a role. The results of the previous studies suffered from the inability 
to distinguish real opacity from foreground confusion as the reason for the 
decrease in field galaxy numbers. Therefore, \cite{Gonzalez98} introduced 
the ``Synthetic Field Method'' (SFM) to calibrate the number of distant galaxies 
for crowding and confusion resulting from the foreground disk and applied it 
to two galaxies.
\cite{Gonzalez03} and \cite{Gonzalez04} explore the limitations of this method 
imposed by the characteristics of the foreground disk: surface brightness, 
granularity and large-scale structure. 

In recent papers in this series \citep{Holwerda05a,Holwerda05b}
\footnote{\cite{Holwerda05a,Holwerda05b} and early versions of \cite{Holwerda05c,Holwerda05e} 
and this paper are presented in \cite{Holwerda05}}, 
we have automated the SFM and analysed a large set of fields of spiral 
galaxies. In this paper we study the limitations of the SFM using this dataset, 
as it spans a range in foreground disk characteristics.

The organisation of this paper is as follows: in section \ref{secLimSFM} the 
SFM is briefly described, section \ref{secLimpredic} describes the predictions 
of \cite{Gonzalez03} relevant to this paper and section \ref{secLimcomp} the 
data from \cite{Holwerda05b} used. In section \ref{secLimSB} we discuss the 
dependence of the SFM on surface brightness and in section \ref{secLimwfpc2} 
the effects of distance, granularity and structure in the foreground disk. Section 
\ref{secLimDisc} discusses the optimum distance for WFPC2 imaging. 
In section \ref{secLimconcl} the conclusions are listed and in section 
\ref{secLimFut} the possibilities for future work are reviewed.

\begin{figure}
\includegraphics[width=0.5\textwidth]{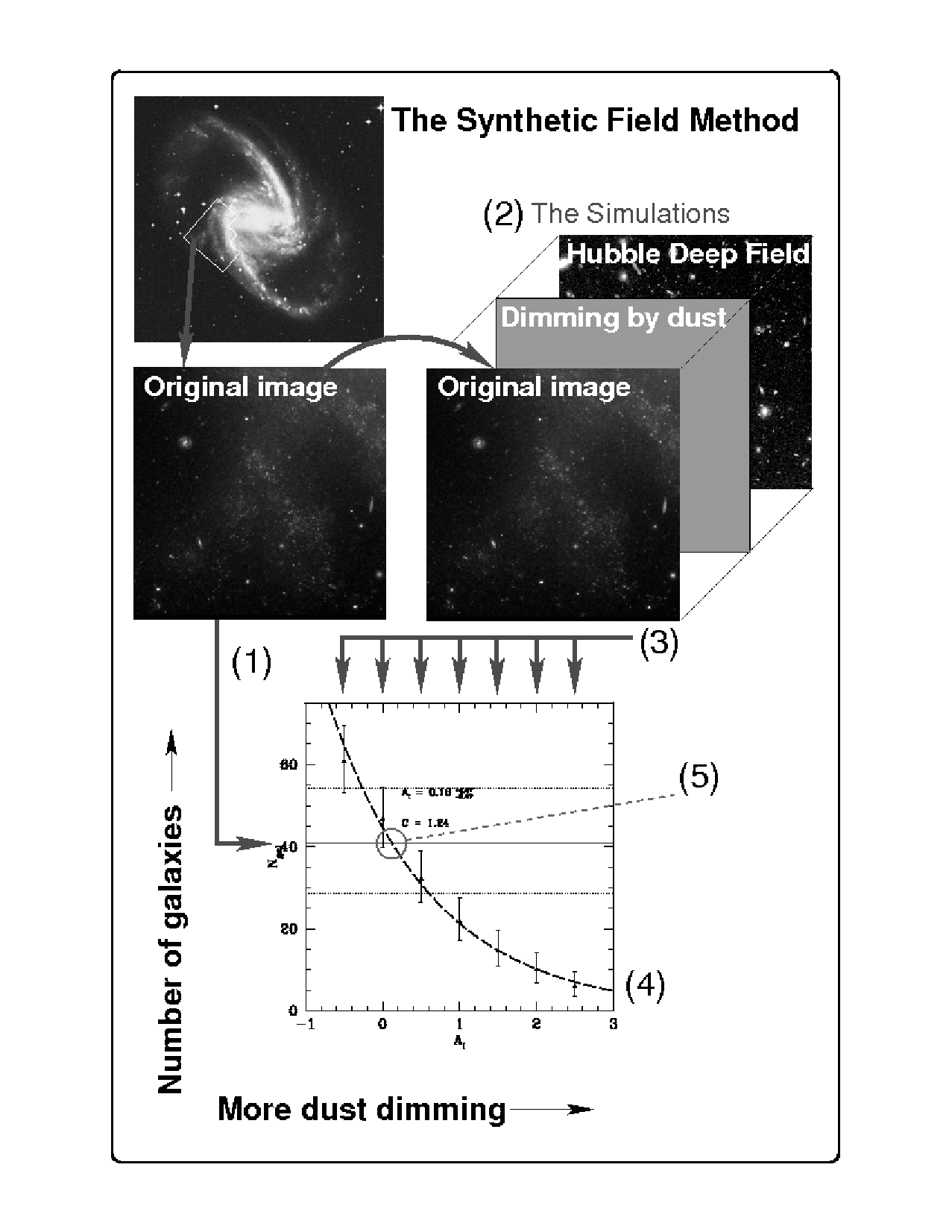}
\caption{A schematic of the ``Synthetic Field Method''. First a WFPC2 field is retrieved 
from the Hubble Space Telescope archive and redrizzled. \protect\\ 
The Synthetic Field Method itself consists of the following steps:\protect\\ 
1. The number of distant galaxies in the original science field are counted. \protect\\ 
2. The ``synthetic fields'' are made by combining a dimmed Hubble Deep Field 
with the science field.\protect\\ 
3. The numbers of synthetic galaxies are counted in the synthetic fields.\protect\\ 
4. Equation \ref{eqNA} it fitted to the number of synthetic galaxies as a function 
of the applied dimming.\protect\\ 
5. From the intersection between the number galaxies in the science field and 
the fit, the average dimming in the image is found.}
\label{method}
\end{figure}

\section{\label{secLimSFM}The ``Synthetic Field Method''}

The number of distant galaxies found in a given field in a spiral disk is indicative 
of the average dust extinction of that field ($A$) but it also depends on the crowding 
and confusion conditions of the field.
The ``Synthetic Field Method'' calibrates the number of distant 
galaxies found in the science field with a series of synthetic fields (See Figure 
\ref{method} for a schematic.). These are the original science field with a Hubble 
Deep Field (North or South) added, which is dimmed to simulate dust extinction. 
In these synthetic fields, the crowding and confusion effects for the detections of 
synthetic galaxies are the same for the distant galaxies in the science field.
Several synthetic fields are made for each value of the dimming.

Each set of synthetic fields is characterised by the applied dimming and 
the average number of synthetic galaxies retrieved for this dimming.
We fit the following relation to the dimming ($A$) of each set and average 
number of galaxies ($N$) retrieved from these sets:

\begin{equation}
\label{eqNA}
A = -2.5 \ C\ log \left({N \over N_0 }\right) 
\end{equation}
 
\begin{figure}
\centering
\includegraphics[width=0.5\textwidth]{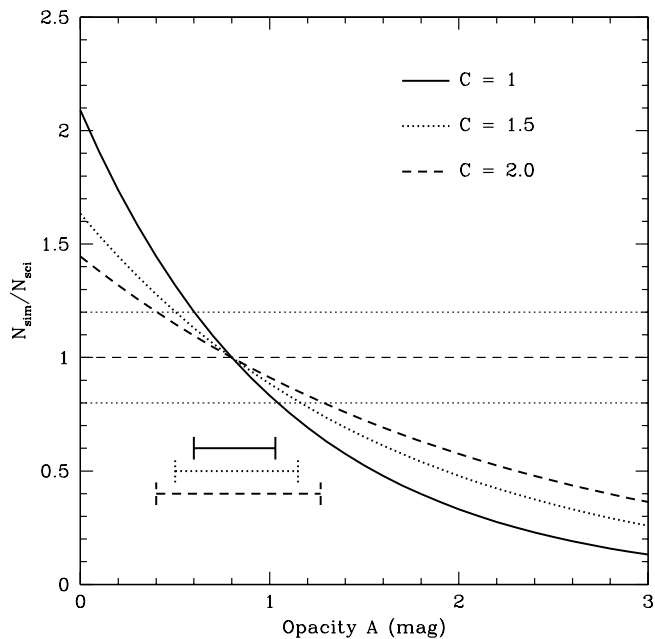}
\caption{\label{N_A_fit}The relation between opacity (A) and the ratio of distant 
field galaxy numbers from the simulations ($N_{sim}$) and science field ($N_{sci}$). 
In the case where $N_{sim} = N_{sci}$, the dimming applied to the simulation is the 
same as the average opacity of the science field. The slope of the relation between 
simulated galaxies and dimming is given by $A ~ = ~ -2.5 ~ C ~ log(N/N_0)$ . Higher 
values of $C$ are caused by surface brightness and crowding effects in the field. 
The uncertainty in the opacity measurement, denoted by the horizontal bars, 
increases with $C$.}
\end{figure}

\noindent In this relation, $C$ is the slope of the relation and $N_0$ the normalization 
(Figure \ref{N_A_fit}). Replacing N with the number of galaxies from the science field 
in equation \ref{eqNA} gives us the average extinction in the field due to dust in the 
foreground disk ($A$).

The normalization ($N_0$) is the number of distant galaxies in the case of no 
dimming. This value depends on the solid angle over which the measurement 
is made and the conditions in the field. 
In the ideal case, the slope is unity ($C =  1$) and the distant galaxy number 
(which can be thought of as a flux) is only reduced due to dimming by dust. However, 
other factors, such as the surface brightness and the crowding of the 
foreground field, influence the detection of distant galaxies. For this 
reason, separate synthetic fields are made and the above relation is fitted for 
each unique science field. 

Uncertainties arise from measurement uncertainties (Poisson statistics) and 
the natural clustering of field galaxies. The latter uncertainty, due to cosmic 
variance in the number of distant galaxies in the science field, can be 
accounted for, as this behaviour is  described by the 2-p correlation function. 
For a more detailed discussion on the uncertainties and systematics, see 
\cite{Holwerda05a} and \cite{Holwerda05}.

The effects of adding foreground objects are twofold: the number of field 
galaxies that can be detected in the field ($N_0$) drops. And secondly, 
the relation between synthetic galaxies and dimming (equation \ref{eqNA}) becomes 
shallower ($C > 1$), as only brighter galaxies distinguish themselves from 
foreground objects. Both these effects result in a more inaccurate determination 
of the average opacity (Figure \ref{N_A_fit}).
The normalisation\footnote{The actual field behind a foreground disk is uncertain 
due to the clustering of distant field galaxies. However, $N_0$ is determined from 
the average of HDF-N/S, a {\it known} background which reasonably approximates 
the average count of galaxies in the sky. Any variations of $N_0$ are therefore the 
result of addition of the synthetic background to a foreground field.} ($N_0$) and 
the slope ($C$) of equation \ref{eqNA}, together with the limiting magnitude of the distant 
galaxies found in the $A = 0$ simulations, are our diagnostics for how well a field is 
suited for the SFM.
 
\section{\label{secLimpredic}Predicted limitations using Hubble}

The detection of distant field galaxies through a foreground disk depends on 
three parameters: (1) the surface brightness of that disk, (2) its granularity, and 
(3) its structure (e.g. spiral arms). \cite{Gonzalez03} measured the effects of the 
foreground disk and instrument resolution on the observable numbers of field 
galaxies.

They divided each field into sections of $100^2$ pixels. For each section the 
mean and standard deviation of the pixel-values are determined.
The average of these mean values is the indicator of surface brightness of the 
field, the average of these standard deviations is the measure for granularity in 
the field and the FWHM of the distribution of mean pixel-values, the indicator of 
large structure. Data and simulations were analysed for the LMC, M31 and 
NGC 4536, probing different disk parameters, distances, and instrument resolution.
\cite{Gonzalez03} parameterised the dependence of distant galaxy detection 
on distance and resolution as follows:

\begin{equation}
\label{eqGonz03}
S/N = {L f_{bg} \over \sqrt { {L^2 n f^2_{*} \over d^2} + n^2 f_{*}d^2}}
\end{equation}

\noindent where $f_{bg}$ and $f_{*}$ are the flux from the distant galaxy and a 
typical disk star respectively, $n$ the number of stars per pixel, $d$ the distance 
of the foreground disk and $L$ the pixel size of various instruments. Whereas 
\cite{Gonzalez03} were interested in varying $L$ at three fixed distances, we 
will explore the effects of varying $d$ and $n$ (granularity and surface brightness) 
at fixed $L$.

\cite{Gonzalez03} concluded that spiral galaxies at the distance of the Virgo 
cluster would make much better candidates for the application of the SFM than 
local group galaxies, and that improvements in resolution would benefit nearby 
foreground galaxies the most.

\section{\label{secLimcomp}Comparison data and systematics}

\cite{Holwerda05b} analysed a sample of 32 WFPC2 fields and presented 
radial opacity plots for both individual galaxies, as well as for the entire 
sample combined. 
In addition, the SFM opacity can be measured for each WFPC2 field as a 
whole\footnote{The Planetary Camera part of the WFPC2 array is not used 
in the SFM analysis. It has different noise characteristics, smaller FOV and 
fewer reference fields.} to characterise the effects of distance of the 
foreground disk. Surface brightness, granularity and structure are 
characterized in the same way as \cite{Gonzalez03}. 

There are two possible sources of systematics for this sample of WFPC2 fields: 
the differences in exposure times and the resampling to 0\farcs05 pixels 
using the ``drizzle'' routine. 

The total exposure time of the images could conceivably influence the 
granularity measure of images if it is the dominant factor in the pixel-to-pixel 
variations\footnote{\cite{Gonzalez98} found that any exposure time above 
2000 seconds did not limit the SFM measurement. \cite{Gonzalez03} concluded 
that the granularity was the predominant limiting factor. Most fields have exposure 
times above 2000 seconds (See Table 3 in \cite{Holwerda05b}).}. The weight 
image from the drizzle routine indicates the relative exposure of pixels in the 
final drizzled image. We compared the pixel-to-pixel variation or the drizzle-
weight image to the total exposure time and found no correlation. 
The fields can therefore be treated as uniform in noise characteristics in the following.

In order to check the effect of image resolution on the detection of distant 
galaxies, the SFM analysis needs to be carried out on the same field with 
different spatial resolutions. \cite{Holwerda05a} compared the numbers of 
distant galaxies in undrizzled WF data from \cite{Gonzalez98} (1 pixel = 0\farcs1) 
with those from their drizzled data (1 pixel = 0\farcs05) of NGC 4536. The 
difference in synthetic galaxy numbers can be attributed to a difference 
between the manual and automated detection methods. 
As \cite{Gonzalez03} predicted, the galaxy statistics were not 
improved at this distance by smaller pixels. However, as it does facilitate 
automated classification, our fields were sampled to the 0\farcs05 scale.

The data from \cite{Holwerda05b} is sufficiently uniform and the resampling 
to a smaller pixel scale will not influence the comparison to the prediction of 
\cite{Gonzalez03}.

\begin{figure}
\centering
\includegraphics[width=0.5\textwidth]{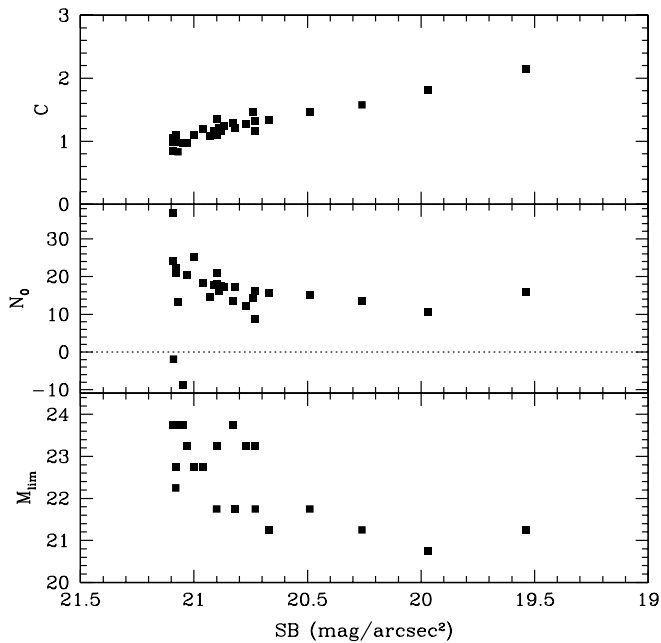}
\caption{\label{SB_C_N0_Mlim}The relation between surface brightness 
in I (F814W) and the $C$ and $N_0$ (in galaxies per square arcmin) 
parameters from equation \ref{eqNA} and the limiting magnitude of the detected 
synthetic field galaxies with no dimming applied. Limiting magnitude 
estimates for the higher surface brightnesses become increasingly 
hindered by poor statistics. (The spread in $N_0$ at lower surface 
brightnesses is from lack of solid angle at those radii, a selection 
effect in the sample of WFPC2 pointings used.) }
\end{figure}

\section{\label{secLimSB}The effects of surface brightness}

\cite{Gonzalez03} briefly illustrated the effect of surface brightness on the SFM's 
accuracy in their figure 7, with a radial sequence for a simulation of M31.
\cite{Holwerda05b} presented average radial opacities based on the counts in 
radial bins, scaled with the $R_{25}$ \citep{RC3}, {\it combined over all fields}.

First, the effects of the surface brightness averaged over the entire sample of 
\cite{Holwerda05b} are shown per radial annulus. Figure \ref{SB_C_N0_Mlim} 
shows the relation between average surface brightness of the radial annuli versus 
the limiting magnitude ($M_{lim}$) of background galaxy detection, and the slope 
($C$) and normalization ($N_0$) in equation \ref{eqNA}. A brighter foreground field is 
expected to limit the magnitude at which distant galaxies can be identified, 
thus limiting the number available for the SFM. If the effect of surface brightness 
dominates the loss of background galaxies, the extinction becomes a secondary 
effect, flattening the slope in equation \ref{eqNA}. 

From Figure \ref{SB_C_N0_Mlim}, it is evident that indeed the surface brightness 
influences the limiting magnitude and hence the accuracy of the SFM
\footnote{The limiting magnitude is estimated in increments of 0.25 mag, hence 
the discrete values in Figures \ref{SB_C_N0_Mlim}, \ref{D_C_N0_Mlim} and 
\ref{D_struc_gran_sb}. In the brightest regions, the limiting 
magnitude estimate becomes uncertain due to the poor statistics.}. Its effect 
on the normalization ($N_0$) of equation \ref{eqNA} is visible. A tight relation between 
average surface brightness and the slope ($C$) is especially evident. 

From the relation between $C$ and surface brightness, it is immediately clear 
that the inner, brighter regions of spiral disks will not ever yield useful opacity 
measurements.
With the effect of surface brightness on $C$ and $N_0$ characterized, it is possible 
to measure opacity without any synthetic fields and derive it directly from the 
number of field galaxies and the average surface brightness of the science field. 
However, the detection method and data-type (WFPC2 field) should be kept the 
same if one is to forgo the synthetic fields completely.


\section{\label{secLimwfpc2}The effects on individual WFPC2 fields}

\begin{figure}
\centering
\includegraphics[width=0.5\textwidth]{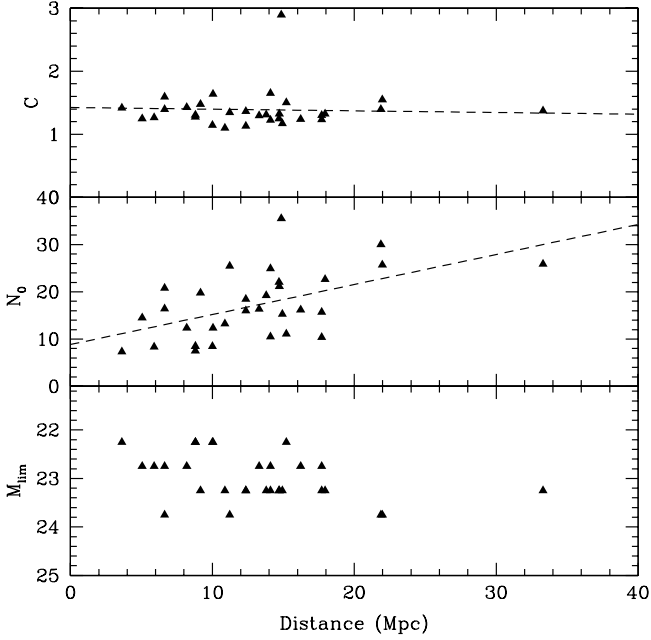}
\caption{\label{D_C_N0_Mlim}The dependence of limiting magnitude 
($M_{lim}$), normalisation ($N_0$, in galaxies per square arcmin) and 
slope ($C$) on distance of the foreground disk. Solid angle effects are 
taken out as each of these points are from one set of three WF chips in 
a WFPC2 mosaic. The dashed lines are linear fits to the points shown.}
\end{figure}

In the previous section, the effects of average surface brightness in the radial 
annuli of the combined fields \citep{Holwerda05b} were discussed. To determine 
the effects of distance and granularity, surface brightness and structure in the 
individual fields, the opacity for the entire WFPC2 for {\it each} foreground galaxy 
was estimated from equation \ref{eqNA} \footnote{In the figure \ref{D_C_N0_Mlim} through \ref{struct_C_N0_Mlim}, triangles are measurements of individual WFPC2 fields 
as opposed to figure \ref{SB_C_N0_Mlim} in which the squares represent 
measurements for the combined radial annuli.}.

\subsection{\label{ssecLimDist}Distance}

The effect of distance of the foreground galaxy on the SFM parameters 
($C,~ N_0,~ M_{lim} $) is plotted in Figure \ref{D_C_N0_Mlim}. Only the normalization 
($N_0$) shows some dependence on distance. As all these points are for the 
combined WFPC2 array, the solid angle is the same for each point. The rise of 
$N_0$ with distance is consistent with the prediction of \cite{Gonzalez03}. The 
granularity of the foreground disk is expected to drop with distance as the foreground disk is smoothed with distance. This allows more background galaxies to be detected in a field. 
The slope C is practically constant with distance. 

\begin{figure}
\centering
\includegraphics[width=0.5\textwidth]{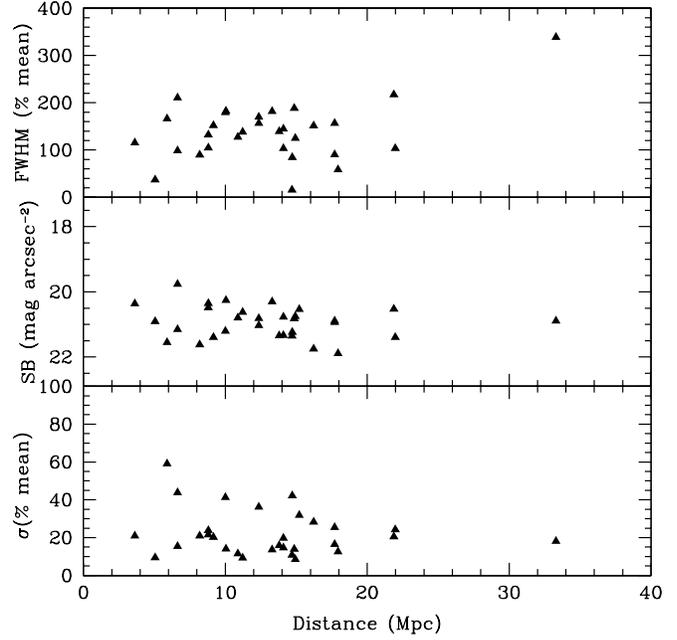}
\caption{\label{D_struc_gran_sb}The dependence of structure, granularity 
and surface brightness of the WFPC2 fields on distance of the foreground 
disk. The top panel shows the FWHM of the distribution of mean values of 
the $100^2$ sections, the middle panel shows the surface brightness 
derived from the average of that distribution. The lower panel shows the 
granularity, the mean of the distribution of the standard deviation of 
pixel-to-pixel variations in each section.}
\end{figure}

\begin{figure}
\centering
\includegraphics[width=0.5\textwidth]{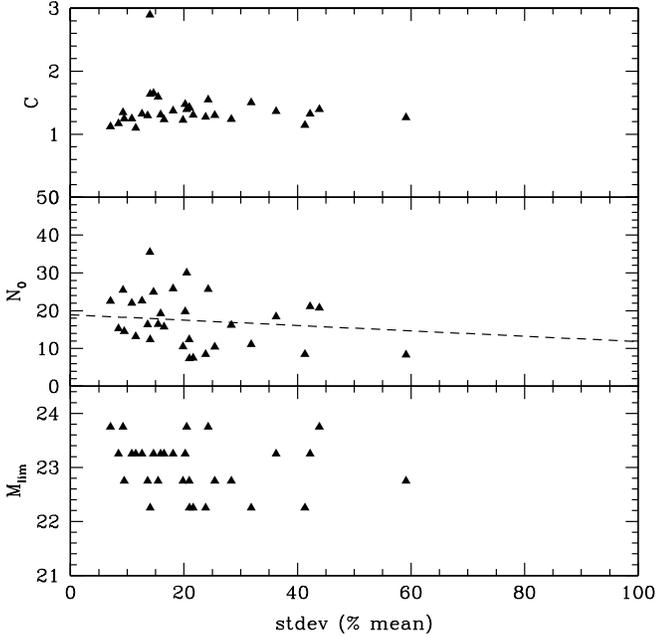}
\caption{\label{gran_C_N0_Mlim}The dependence of limiting magnitude 
($M_{lim}$), normalisation ($N_0$, in galaxies per square arcmin) and 
slope ($C$) on granularity of the foreground disk. Granularity is characterized 
by the mean $\sigma$ as a percentage of the mean pixel value of the 
$100^2$ pixel sections. Only $N_0$ seems to decline with granularity. 
The dashed line is a linear fit to those points.} 
\end{figure}

To check that granularity is the cause for this trend with distance, we plot the 
relations between structure, surface brightness and granularity with distance in 
Figure \ref{D_struc_gran_sb}. Surface brightness is not expected to change 
with distance, therefore this serves as a check against a systematic selection 
effect in our fields which could influence the granularity result.

The spread in granularity (i.e. $\sigma$) does seem to drop with distance while
surface brightness and structure (FWHM) do not change much with distance.

\subsection{\label{ssecLimGran}Granularity}

Figure \ref{gran_C_N0_Mlim} shows the direct effect of granularity on the SFM 
parameters ($C,~ N_0,~ M_{lim} $) and the effect of distance on $N_0$ seen 
in Figure \ref{D_C_N0_Mlim} appears due to the granularity effect of smoothing 
the foreground disk with distance. While a some trend for $N_0$ with distance and 
granularity can be distinguished, it is not tight enough to forgo a synthetic 
field to characterize $N_0$ altogether. However, as it is a quick diagnostic, candidate fields can be 
ranked in order of expected SFM accuracy based on this relation. The relation 
between normalization $N_0$ and distance (Figure \ref{D_C_N0_Mlim}) is 
expected to level out at the number of distant galaxies which are relatively 
easily identified in the Hubble Deep Fields, about 30 galaxies per square 
arcminute. Accordingly, at small granularity ($\sigma$), the relation between 
$N_0$ and granularity (Figure \ref{gran_C_N0_Mlim}), reaches that same number.

A WFPC2 field of a disk beyond 15 Mpc. has a factor 2-3 more identifiable 
distant galaxies in the $A = 0$ reference field than one of a closer disk 
($\rm d < 10 ~ Mpc$, see Figure \ref{D_C_N0_Mlim}). Therefore, the strategy 
of \cite{Holwerda05b} to combine numbers from fields at greater distances 
maximizes the detection of distant galaxies and hence the accuracy of the 
method. Increasing solid angle on a single nearby foreground disk has less 
efficiency than adding solid angle to these, more distant, disks. 

\begin{figure}
\centering
\includegraphics[width=0.5\textwidth]{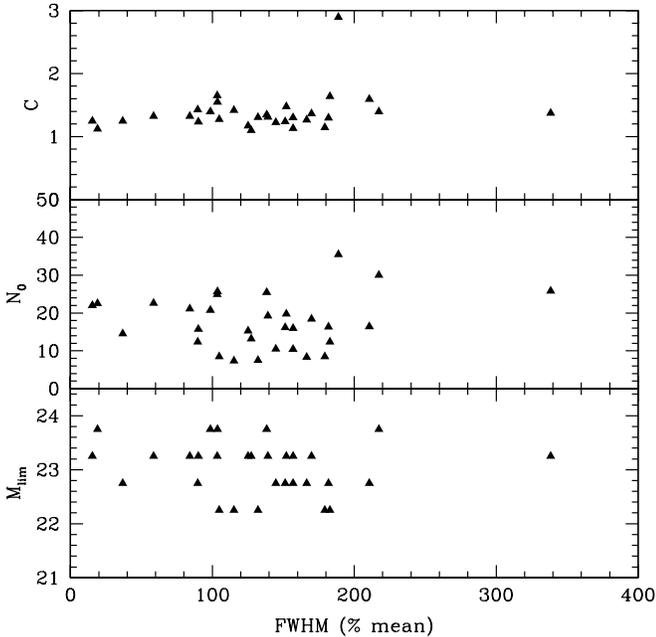}
\caption{\label{struct_C_N0_Mlim}The dependence of limiting magnitude 
($M_{lim}$), normalisation ($N_0$, in galaxies per square arcmin) and 
slope ($C$) on structure in the foreground disk. Strucure is characterized 
by the FWHM of the distribution of mean pixelvalues of the $100^2$ pixel 
sections. Expressed as a percentage of the mean of that distribution. }
\end{figure}

\subsection{\label{ssecLimStruct}Effects of structure and surface brightness}

Structure in a field can be of importance for the application of the SFM. 
A spiral arm raises the surface brightness and adds to the crowding and 
confusion. In Figure \ref{struct_C_N0_Mlim} the relation between the structure 
(FWHM of the distribution of the mean pixel values of the image sections) and 
the SFM parameters is shown. 
Structure shows little effect on the SFM parameters, except for an effect on C similar to 
the surface brightness. 
%
The average surface brightness of the WFPC2 field has little effect on the 
SFM parameters as most of the flux can be from one section of the field 
while the SFM measurement is done in another. A spread of surface 
brightness values over a field will introduce a spread in the relation with $C$ 
in Figure \ref{SB_C_N0_Mlim}. 


\section{\label{secLimDisc}Discussion: Optimum Distance for the SFM}

The optimum distance range for the SFM applied on HST imaging (WFPC2 
and ACS) is limited by two factors, the solid angle covered by (part of) the 
foreground disk for which an opacity measurement needs to be made and 
the granularity of the foreground disk. The granularity imposes a minimum 
distance, the solid angle a maximum. The solid angle does not only depend 
on the distance but also on the intrinsic size of the foreground galaxy. In 
addition to that, not all of the disk is suitable for SFM opacity measurements. 
We consider M101 as a template face-on spiral galaxy ($R_{25} = 28.1$ 
kpc., D = 6.7 Mpc) with the inner 25\% of that radius unsuitable for SFM 
measurements due to high surface brightness.

The {\it maximum} distance to which the SFM can be used, is determined by 
the minimal statistics -and hence solid angle- for which a opacity measurement 
can still be performed. To illustrate this, we assume that the minimal 
measurement is one magnitude of opacity, measured over the entire disk 
with an accuracy of $\pm0.75$ mag. 

An estimate of the error in the opacity measure needs an expression for 
the uncertainties in the number of galaxies. One can approximate the error in 
the number of surviving galaxies in the science field as $\Delta N = \sqrt{2N}$. 
This is an overly simplistic but useful analytic approach to the clustering uncertainty. 
The uncertainty in the numbers of synthetic field galaxies is a simple Poisson 
uncertainty: $\Delta N_0 = \sqrt{N_0}$.
Thus, we get the following expression for the uncertainty in opacity $A$:

\begin{equation}
\label{eqErr}
\Delta A  = 2.5 ~ C ~ \sqrt{ {2 \over N} + {1 \over N_0}  }
\end{equation}

\noindent Typical SFM parameters for a disk are $\rm C = 1.2$, $\rm 
N_0 = 25 ~ gal ~ arcmin^{-2}$, $\rm M_{lim} \approx 23 ~ mag.$ (Figure 
\ref{D_C_N0_Mlim}). An opacity of 1 magnitude would result in $\rm 
N_{science} = 11.6$ field galaxies per square arcminute (equation \ref{eqNA}). 
To fulfil the requirement of $\rm \Delta A = 0.75$, we need the disk to 
cover 3.39 square arcminutes for a meaningful measurement. If we 
assume that the inner 25 \% of the disk is too bright for use, the 
maximum distance becomes approximately 70 Mpc for a disk the size 
of M101. 

However, M101 is a large galaxy and most galaxies are not that neatly 
face-on. The effective maximum distance will therefore be much less 
for a single disk\footnote{Many disks could in principle be combined to 
improve statistics.}. Also, a measurement of extinction $\rm A = 1 \pm 0.75$ 
for the entire disk hardly warrants the effort. For individual galaxies a 
maximum distance of 35 Mpc should be considered much more 
practical, allowing for some spatial resolution of opacity in the disk.

The {\it minimum} distance also depends on the choice of solid angle of 
interest, in addition to the effects of granularity. If the measurements are 
taken over larger solid angles, the loss of distant galaxies due to 
granularity can be compensated for. The relation between the number of 
field galaxies without extinction ($N_0$) as a function of distance is shown 
in Figure \ref{D_C_N0_Mlim}. The number of field galaxies that can be 
detected through the disk decreases by a factor 2-3 when the distance drops 
from around 25 Mpc down to below 10. If we still want to get an opacity 
measurement similar to the one above ($\rm A = 1, \Delta A = 0.75$) 
then the minimum solid angle required is $3.39 \times 2,5 \approx 8.4$ 
square arcminutes.

In order to compensate for the loss due to granularity, a bigger solid angle 
can be considered, which is possible for a closer disk. However another 
consideration comes into play: the field-of-view of the instrument. Even 
with the $12.25 $ square arcminutes of the ACS, the maximum surveyed 
solid angle for a single galaxy is some $180$ square arcminutes. (Two 
recent programs on M101, not covering the whole of the disk).
So the biggest solid angle surveyed on a single galaxy to date has about 
20 SFM resolution elements in it. A galaxy disk at a shorter 
distance would require an even bigger investment of observing time. 
This makes the distance to M101 of 6.7 Mpc the minimum practical distance. 
This distance can be brought down some by sacrificing some spatial 
resolution. But the minimal value for $N_0$ in Figure \ref{D_C_N0_Mlim} 
imply marginal results.

A larger solid angle is easier to cover using ground-based observations. 
As \cite{Cuillandre01} showed for M31, the confusion between blended 
foreground stars and background galaxies quickly makes a meaningful 
measurement impossible, even at some distance from the galaxy center. 
The imaging standards of the SFM effectively demand space-based optical data.

The prediction from \cite{Gonzalez03} that the SFM can only successfully 
be applied on Virgo cluster spiral galaxies is corroborated by these 
estimates of minimum and maximum distance. Opacity measurements of 
Local Group members will indeed be much more difficult. 

\section{\label{secLimconcl}Conclusions}

From the uniform application of the ``Synthetic Field Method'' to a sample 
of HST/WFPC2 fields, we draw some conclusions as to the applicability of 
this method on HST images of spiral disks:\\
\begin{itemize}
\item[1.] Surface brightness affects the accuracy of the SFM by flattening the slope 
in equation \ref{eqNA} (Figure \ref{SB_C_N0_Mlim}). The relation between SB and C 
is remarkably without much scatter. This relation limits the SFM to the outer 
regions of the foreground disk.
\item[2.] Granularity affects the accuracy by diminishing the detectability of 
galaxies and hence the normalization of equation \ref{eqNA} (Figure \ref{gran_C_N0_Mlim}).
\item[3.] There is a downward trend of granularity with distance. This is consistent with the 
prediction from \cite{Gonzalez03} that this is the limiting factor for nearby disks.
(Figure \ref{D_C_N0_Mlim} and \ref{D_struc_gran_sb}).
\item[4.] Surface brightness averaged over a field and structure in a field have 
a similar effect on the SFM. They limit its accuracy in the center of the disk 
(Figures \ref{SB_C_N0_Mlim} and \ref{struct_C_N0_Mlim}).
\item[5.] The effective minimum distance for the SFM would be of interest 
as its use on nearby galaxies could give us the most detailed opacity map 
of a disk. Using reasonable numbers, a minimum distance of 5 Mpc is found 
from the relations between SFM parameters and distance, due to granularity 
and FOV considerations.
\item[6.] The effect of a foreground disk on the number of distant galaxies can 
be detected up to some 70 Mpc but the effective maximum distance for any 
scientific interesting result is about 35 Mpc. This would provide some spatial resolution 
of the dust extinction. 
\item[7.] The relation between granularity and SFM accuracy displays still 
some scatter. Hence a synthetic field to chatacterize the normalisation 
($N_0$) may be desirable. 
However, a quick result can be immediately obtained from the field's 
characteristics, surface brightness and granularity and the number of 
distant galaxies detected, provided detection method and data are similar to this paper's.
These relations should also help in the selection of ACS fields for SFM analysis.
\item[8.] Future work with the ACS seems more than feasible, even for the 
closer disks. The combination of its resolution, sensitivity and field-of-view 
will likely facilitate measurements. The FOV and speed tips the balance 
more in favour of nearer objects.
\end{itemize}

\section{\label{secLimFut}Future work}

With the SFM proven to function, the number counts for other HST imaging 
of face-on spiral galaxies could be used for opacity measurements. The 
Advanced Camera for Surveys has superior field-of-view and sensitivity 
making its fields of face-on spirals obvious candidates. Currently in the 
Hubble archive are fields of NGC300, NGC3370, NGC3621, NGC3949, 
NGC4258, NGC4319 and notably large datasets on M51 and M101. 
These span a range of distances, NGC3370 near the possible maximum 
(D = 30 Mpc) and NGC300 below the minimum (D = 2 Mpc). These 
however are imaged in more photometric bands, improving the field 
galaxy identification, and the Hubble Ultra-Deep Field \citep{UDF} and 
GOODS fields are candidate reference fields. With this wealth in existing 
data, the SFM promises to continue to shed light on dust extinction.

\acknowledgements

The authors would like to thank the anonymous referee for his or her comments.
This research has made use of the NASA/IPAC Extragalactic Database, 
which is operated by the Jet Propulsion Laboratory, California Institute of 
Technology, under contract with the National Aeronautics and Space 
Administration (NASA). This work is based on observations with the 
NASA/ESA Hubble Space Telescope, obtained at the STScI, which is 
operated by the Association of Universities for Research in Astronomy 
(AURA), Inc., under NASA contract NAS5-26555. Support for this work 
was provided by NASA through grant number HST-AR-08360 from
STScI. STScI is operated by AURA, Inc., under NASA contract
NAS5-26555. We are also grateful for the financial support
of the STScI DirectorÕs Discretionary Fund (grants 82206 and
82304 to R. J. Allen) and of the Kapteyn Astronomical Institute 
of the University of Groningen.


\end{document}